\documentclass[prl,aps,twocolumn,preprintnumbers,showpacs,superscriptaddress]{revtex4}
\usepackage{latexsym,epsfig,amssymb,amsfonts,amsmath,graphicx}

\newcommand{\ket}[1]{\left | \, #1 \right \rangle}
\newcommand{\bra}[1]{\left \langle #1 \, \right |}
\newcommand{\braket}[2]{\left\langle\, #1\,|\,#2\,\right\rangle}

\newcommand{\tr}[1]{\mbox{Tr} \, #1 }

\def\id{\mathbb{I}}

\allowdisplaybreaks

\begin{document}

\title{Multipartite entanglement detection in bosons}

\author{C. \surname{Moura Alves}}\email{carolina.mouraalves@qubit.org}%
\affiliation{Clarendon Laboratory, University of Oxford, Parks Road, Oxford OX1 3PU , U.K.}%
\affiliation{Centre for Quantum Computation,
DAMTP, University of Cambridge, Wilberforce Road, Cambridge CB3 0WA , U.K.}%
\author{D. \surname{Jaksch}}%
\affiliation{Clarendon Laboratory, University of Oxford, Parks Road, Oxford OX1 3PU , U.K.}%

\date{\today}

\begin{abstract}
We propose a simple quantum network to detect multipartite
entangled states of bosons, and show how to implement this network
for neutral atoms stored in an optical lattice. We investigate the
special properties of cluster states, multipartite entangled
states and superpositions of distinct macroscopic quantum states
that can be identified by the network.
\end{abstract}


\maketitle

Quantum entanglement is a very important physical resource in
quantum information processing tasks, such as quantum
cryptography, teleportation, quantum frequency standards or
quantum-enhanced positioning~\cite{Entanglement}. Since the
implementation of any of these tasks requires precise knowledge on
the entangled states being used, the development of ``measurement
tools" for the characterization and detection of entanglement in
physical systems is of great practical importance.

While the creation of multipartite entanglement has been recently
achieved in a controlled way in experiments with either Mott
insulating states of neutral bosonic atoms in optical lattices
\cite{BlochEnt,BECinOL} or ion traps \cite{ClassQuant}, its
unambiguous detection has so far not been possible \cite{BlochEnt}
in these setups. The usual experimental methods to detect
entanglement are based on the violation of Bell type inequalities
\cite{Bell64}, which are known to be quite inefficient, in the
sense that they leave many entangled states undetected
\cite{Werner}. Alternatively, one can perform a complete state
tomography of the system \cite{Tomography}, but this method
requires the preparation of an exponentially large number of
copies of the state and it is redundant, since not all parameters
of the density operator are relevant for the entanglement
detection.

\begin{figure}[tbp]
\includegraphics[width=8.5cm]{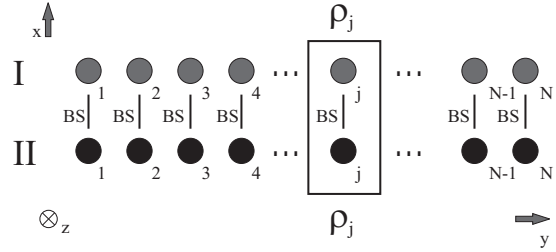}
\caption{Network of BS acting on pairs of identical bosons. The
two rows of $N$ atoms, labelled $I$ and $II$ respectively, are
identical, and the state of each of the rows is $\rho_{123...N}$.
The total state of the system is $\rho_{123...N} \otimes
\rho_{123...N}$.} \label{Fig1}
\end{figure}

In this Letter we propose a simple quantum network to detect
multipartite entangled states through an entanglement test more
powerful than the Bell-CHSH inequalities for all possible settings
\cite{Entropies}, albeit less powerful than full state tomography.
The network is realized by coupling two identically prepared 1D
rows of $N$ previously entangled qubits via pairwise beam
splitters (BS), as shown in Fig.~\ref{Fig1}. We also show how to
implement this network in an optical lattice or array of magnetic
microtraps loaded with atoms in a Mott insulating state with
filling factor one \cite{Jaksch98}. Each of the atoms has two long
lived internal states $a$ and $b$ which represent the qubit. The
pairwise BS can be implemented by decreasing the horizontal
barrier between the two rows of atoms.

We start by introducing the set of inequalities used by our
network for the detection of multipartite entanglement. The
information-theoretic approach to separability of bipartite
quantum systems leads to a set of entropic inequalities satisfied
by all separable bipartite states \cite{Entropies}. We extend
these inequalities to separable multipartite states by considering
a state $\rho_{123...N}$ of $N$ subsystems. If $\rho_{123...N}$ is
separable then we can write it as
\begin{eqnarray}
\rho_{123...N}= \sum_{\ell} C_\ell \rho^{\ell}_1 \otimes
\rho^{\ell}_2 \otimes \rho^{\ell}_3 \otimes \ldots \otimes
\rho^{\ell}_N, \label{rho}
\end{eqnarray}
where $\rho^{\ell}_j$ is a state of subsystem $j$, and
$\sum_{\ell} C_\ell=1$. The purity $\tr(\rho_{123...n}^2)$ of
$\rho_{123...n}$, where $n\in 1,2,...,N$, is smaller or equal than
the purity of any of its reduced density operators. For example,
\begin{eqnarray}
\tr(\rho_{123...n}^2) &\leq& \tr(\rho_{123...n-1}^2) \leq
\tr(\rho_{123...n-2}^2) \nonumber  \\ &\ldots&  \leq
\tr(\rho_{12}^2) \leq \tr(\rho_{1}^2) \leq 1. \label{ineq}
\end{eqnarray}
This set of nonlinear inequalities provides a set of necessary
conditions for separability, i.e.~if for any state $\rho$ any of
these inequalities is violated then $\rho$ is entangled. For the
case where $\rho_{123...n}$ is separable and pure we have that
$\tr(\rho_{123...n}^2)=1$ and the inequalities become equalities.

In order to test Eq.~(\ref{ineq}) we need to be able to determine
the non-linear functional $\tr(\rho^2)$, where $\rho$ is any of
the different reduced density operators of $\rho_{123...n}$. The
direct estimation of this functional has been addressed in
\cite{Carolina2002-3}, where the value of $\tr(\rho^2)$ is
determined by measuring the expectation value, on state $\rho
\otimes \rho$, of the symmetric and antisymmetric projectors.

Let us first consider the simple scenario of one pair of identical
bosons, in state $\rho_j \otimes \rho_j$, impinging on the BS (as
depicted in Fig.~\ref{Fig1}). After applying the BS we write the
purification of the state $\rho_j'={\cal U}_{\rm BS} \rho_j
\otimes \rho_j {\cal U}_{\rm BS}^\dagger$, where ${\cal U}_{\rm
BS}$ is the unitary time evolution operator of the BS, in the
triplet/singlet basis. The operator ${\cal U}_{\rm BS}$ transforms
$a(b)^{(j)}_{I,II} \rightarrow (a(b)^{(j)}_{I,II} -i
a(b)^{(j)}_{II,I})\sqrt{2}$, where $a_l^{(j)}$ and $b_l^{(j)}$ are
the bosonic destruction operators for particles in row $l=I,II$,
site number $j=1\cdots N$ and internal state $a$ and $b$,
respectively. The triplet basis elements are
$(\alpha^{(j)\dagger}_I \beta^{(j)\dagger}_I +
\alpha^{(j)\dagger}_{II} \beta^{(j)\dagger}_{II})|{\rm
vac}\rangle$, $\alpha,\beta \in {a,b}$, with respective
pre-factors $c_{\alpha\beta}^{(j)}$, and the singlet element is
$(a^{(j)\dagger}_I b^{(j)\dagger}_{II} - a^{(j)\dagger}_{II}
b^{(j)\dagger}_I)|{\rm vac}\rangle$, with pre-factor $c_-^{(j)}$,
and $|{\rm vac}\rangle$ the vacuum state. The
symmetric/antisymmetric components of $\rho_j \otimes \rho_j$ are
spanned by the triplet/singlet basis elements. Note that in the
triplet basis states, both bosons are in the same spatial mode,
while in the singlet basis state each boson occupies a different
spatial mode. Hence, measuring the probability of finding two
bosons in, either the same spatial mode or different spatial
modes, after the BS, is tantamount to measuring the probabilities
$P^i_{\pm}$ of projecting the state $\rho_j \otimes \rho_j$ on its
symmetric or antisymmetric subspaces. These probabilities are
given by the expectation value of the symmetric/antisymmetric
projectors $S_{\pm}$ defined by $S_{\pm}=(\id \pm V)/2$, where
$\id$ is the identity operator, $V$ is the swap operator, with $V
\alpha^{(j)\dagger}_I \beta^{(j)\dagger}_{II} \ket{\textrm{vac}}=
\beta^{(j)\dagger}_I \alpha^{(j)\dagger}_{II} \ket{\textrm{vac}},
\forall \alpha^{(j)\dagger}_I,\beta^{(j)\dagger}_{II}$:

\begin{eqnarray}
P^j_{\pm} &=& \frac{1}{2}\tr[(\id \pm V)\rho_j\otimes\rho_j]=
\frac{1}{2} \pm \frac{1}{2}\tr(\rho_j^{2}),
\end{eqnarray}
where $``-" (``+")$ stands for projection on the (anti)symmetric
subspace. Thus, after we let $\rho_j \otimes \rho_j$ go through
the pairwise BS we can determine the purity of $\rho_j$ by
measuring the projections of $\rho_j'$ on the symmetric and
antisymmetric subspaces. For pairs of polarization entangled
photons, a scenario similar to this one is currently being
implemented experimentally \cite{Nonlin}.

We now extend the above two-boson scenario to the general
situation (see Fig.~\ref{Fig1}) and consider two copies of a state
of $N$ bosons, undergoing pairwise BS. By correlating the
probabilities of projecting the state of each pair of identical
bosons on the symmetric/antisymmetric subspaces, we can estimate
the purity of $\rho_{123...N}$ and of any of its reduced density
operators. As a more concrete example, let us consider the
probabilities for $N=3$. We will label the subsystems $1,2,3$,
respectively:
\begin{eqnarray}
P_{\pm_1\pm_2\pm_3} &=& \frac{1}{2^3}\tr[\prod_{i=1}^3(\id \pm_i
V_i)\rho_{123}\otimes\rho_{123}] \label{prob}\\
&=& \frac{1}{8} \big[ 1 \pm_1 \tr(\rho_1^2) \pm_2 \tr(\rho_2^2)
 \pm_3 \tr(\rho_3^2) \nonumber \\ && \pm_{1,2} \tr(\rho_{12}^2)
\pm_{1,3} \tr(\rho_{13}^2) \pm_{2,3} \tr(\rho_{23}^2) \nonumber
\\ && \pm_{1,2,3} \tr(\rho_{123}^2) \big] \nonumber,
\end{eqnarray}
where $\pm_{i,i'}=(\pm_i)(\pm_i')$, $i,i'=1,2,3$, and $V_{1,2,3}$
stand for the swap operator acting on subsystem $1,2,3$. The
purities related to $\rho_{123}$ are unequivocally determined by
the eight probabilities $P_{\pm_1\pm_2\pm_3}$. The expression for
the probabilities in Eq.~(\ref{prob}) can be straightforwardly
extended to states of $N$ bosons, where we consider the
expectation values of the projector $\prod_{i=1}^N (\id \pm_i
V_i)/2$, on $\rho_{123...N}\otimes\rho_{123...N}$. In the $N$
boson case, the $2^N-1$ unknown purities will be determined by the
$2^N-1$ independent probabilities.

The implementation of this entanglement detection scheme in
optical lattices and magnetic microtraps follows four steps: (i)
Creation of two identical copies of the entangled state
$\rho_{123...N}$: Each of the two rows of bosons shown in
Fig.~\ref{Fig1} is realized by a 1D chain of entangled atoms. The
entanglement can e.g.~be created by spin selective movement and
controlled interactions between atoms as described in
\cite{BlochEnt} or by entangling beam splitters as investigated in
\cite{Dorner}. We assume that any hopping of atoms between the
lattice sites is initially turned off and that the two chains
consist of exactly one atom per lattice site \cite{Jaksch98}. (ii)
Implementation of the pairwise BS: This is achieved by decreasing
the potential barrier between the two rows of atoms. In an optical
lattice one can decrease the corresponding laser intensities
\cite{Jaksch98} while in an array of magnetic microtraps
electric/magnetic fields can be switched to change the barrier
height \cite{Calarco}. The dynamics after lowering the potential
barrier is described by the Hamiltonian $H=H_{\rm BS}+H_{\rm int}$
where (c.f.~\cite{Jaksch98})
\begin{eqnarray}
H_{\rm BS}&=& \sum_{j=1}^N -J (a_I^{(j)\dagger} a_{II}^{(j)} +
b_I^{(j)\dagger} b_{II}^{(j)} + {\rm h.c.} ) \nonumber \\
H_{\rm int} &=& \sum_{l=I,II}\sum_{j=1}^N \frac{U_a}{2}
a_l^{(j)\dagger}a_l^{(j)\dagger} a_{l}^{(j)}a_{l}^{(j)} +
 \nonumber \\ &&
\frac{U_b}{2}
b_l^{(j)\dagger}b_l^{(j)\dagger}b_{l}^{(j)}b_{l}^{(j)} + U_{ab}
b_l^{(j)\dagger}b_{l}^{(j)} a_l^{(j)\dagger}a_{l}^{(j)}.
\label{Hamiltonian}
\end{eqnarray}
Here $H_{\rm BS}$ describes vertical hopping of particles between
the two rows with hopping matrix element $J$ \cite{Pachos03} and
$H_{\rm int}$ gives the on-site interaction of two particles in a
lattice site with internal state dependent interaction strengths
$U_a$, $U_b$ and $U_{ab}$. For simplicity we assume that the
interaction terms $H_{\rm int}$ can be neglected while the hopping
is turned on, i.e.~$J \gg U$, and we assume $U_a =U_b = U_{ab}=U$
\cite{Foot1}. Turning on the Hamiltonian $H_{\rm BS}$ for the
specific time $T= \pi/(4J)$ implements the $N$ pairwise BS. (iii)
Acquisition of a relative phase between the symmetric and
antisymmetric parts of the wave function: After implementing the
BS we let the system evolve according to $H_{\rm int}$ for time
$\tau$. This introduces a phase $\theta=U \tau$ in each doubly
occupied lattice site while it has no influence on singly occupied
lattice sites. Recently it was demonstrated in an interference
experiment \cite{Measure} that this phase $\theta$ allows the
double occupancy sites to be distinguished from single occupancy
ones. (iv) Turning off the lattice and measuring the resulting
interference pattern \cite{Measure}: After time $\tau$ the
particles are released from the trap such that their wave function
dominantly spreads along the vertical direction $x$ (see
Fig.~\ref{Fig1}). The density profile resulting from the pair of
atoms $j$ in state $\rho_j'$ will exhibit interference terms
dependent on $c_{-}^{(j)}$, $c_{\alpha \beta}^{(j)}$ and $\theta$,
so by varying the interaction phase $\theta$, and noting that
$P^j_+=1-P^j_-$, we can determine $|c_-^{(j)}|^2$. For $N$ pairs
of atoms the density profile will depend on $|c_{\alpha
\beta}^{(j)}|^2$ and $|c_{-}^{(j)}|^2$ as well as on correlations
between the density profiles of different pairs of atoms,
according to Eq.~(\ref{prob}). Measuring the density profile for
the $N$ pairs of atoms allows us to determine the different joint
probabilities $P_{\pm\pm...\pm}$, so by solving Eq.~(\ref{prob})
we can detect multipartite entangled states which violate
Eq.~(\ref{ineq}). We note that both the creation of the two copies
of $\rho_{123...N}$ and the network can be implemented with
current experimental technology and do not introduce any novel or
unknown sources of imperfections.

\begin{figure}
\includegraphics[width=8.5cm]{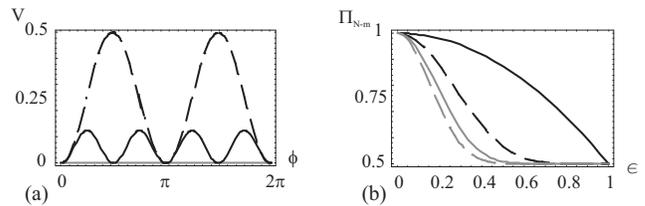}
\caption{In Fig.~2(a), we plot the violation $V$ of the
inequalities Eq.~(\ref{ineq}),
$V_1=\tr(\rho_{123}^2)-\tr(\rho_{12}^2)$ (dashed),
$V_2=\tr(\rho_{12}^2)-\tr(\rho_{1}^2)$ (grey) and
$V_3=\tr(\rho_{12}^2)-\tr(\rho_{2}^2)$ (solid), as a function of
the phase $\phi$, for $N=3$ atoms. Whenever $V>0$, entanglement is
detected by our network. In Fig.2(b), we plot the purity
$\Pi_{N-m}$ for $m=1$ (solid black), $m=7$ (dashed black), $m=14$
(solid grey) and $m=20$ (dashed grey), as a function of
$\epsilon$, for $N=300$ atoms.}\label{Fig2}
\end{figure}

We now focus on the predictions of the entanglement detection
network for multipartite entangled states recently generated in
optical lattices by letting atoms in the chain interact with their
nearest-neighbors \cite{BlochEnt}. The state generated for an
arbitrary number of atoms $N$ is given by
$(1+e^{i\phi})/2\ket{000...0}+ (1-e^{i\phi})/2\ket{\mbox{Cluster
State}}$, where $\phi$ is the phase arising from the controlled
interactions during the entanglement creation procedure and the
precise definition of the cluster state is given in
\cite{Cluster}. The entangled state of $N$ atoms generated by the
process described in \cite{BlochEnt} will violate the inequalities
Eq.~(\ref{ineq}) $\forall N$. For $N=2$ tracing out one atom
yields a violation $V$ of Eq.~(\ref{ineq}) for $\phi \neq \{0,
\pi, 2\pi \}$, i.e.~our network detects all entangled states and
is thus more powerful than Bell-CHSH inequalities which are only
violated for $\phi \in [\varphi, \pi-\varphi]\cup [\pi+\varphi,
2\pi-\varphi]$ with $\varphi \approx 0.7$.
When starting with $N=3$ tracing marginal atoms yields violations
$V$ as shown by the dashed and solid curves in Fig.~\ref{Fig2}(a),
respectively. Again, the network detects and distinguishes all
multipartite entangled states created in this case. Also for $N>3$
multipartite entanglement is detected by tracing the marginal
atoms. However, tracing further atoms does not yield additional
violations of Eq.~(\ref{ineq}) as indicated by the grey line in
Fig.~\ref{Fig2}(a), but the purity of the further reduced states
remains constant. This is due to the characteristic of cluster
states being resilient to measurements on individual subsystems
\cite{Cluster}. In fact, the knowledge of the cluster states'
purities can be further exploited to identify defects (empty
sites) in the lattice and errors in the entanglement creation
process~\cite{Rebecca}.

Let us also note that any pure maximally entangled multipartite
state can unequivocally be identified through its unique property
of being pure while all reduced density operators have purity
$1/2$. In optical lattices such maximally entangled generalized
GHZ states that could be tested with the network can e.g.~be
created by beam splitter setups in 1D atomic pipelines as shown in
\cite{Dorner}.

Our network can also be used to study superpositions of distinct
quantum macroscopic states which are of great importance for the
better understanding of fundamental aspects of quantum
theory~\cite{ClassQuant,Measurement}. There have been several
proposals on how to create macroscopic superpositions in systems
ranging from superconductors \cite{SuperCond Cat}, Bose-Einstein
condensates (BECs) \cite{BEC Cat} to opto-mechanical setups
\cite{OptoMech Cat}. In the case of BECs, the macroscopic
superpositions are multipartite entangled states of the form
\begin{eqnarray}
\ket{\psi}=\frac{1}{\sqrt{K}}(\ket{\phi_1}^{\otimes N} +
\ket{\phi_2}^{\otimes N}) \label{cat},
\end{eqnarray}
where $K=2+ \braket{\phi_1}{\phi_2}^N + \braket{\phi_2}{\phi_1}^N$
and we define a parameter $\epsilon$ by the overlap
$\epsilon^2=1-|\braket{\phi_1}{\phi_2}|^2$. Recently, a measure
based on $\epsilon$ for the effective size $S$ of such
superpositions of distinct macroscopic quantum states was
introduced \cite{Cat}. It compares states of the form $\ket{\psi}$
with generalized GHZ states of $N$ atoms $(\ket{0}^{\otimes N} +
\ket{1}^{\otimes N})/\sqrt{2}$, where $\epsilon=1$ for a
generalized GHZ state. The effective size $S$ of the state
$\ket{\psi}$ is given by $S=N\epsilon^2$ \cite{Cat}.

We can determine $S$ from the measurement of the purity of any
reduced density operator of Eq.~(\ref{cat}). We derive an explicit
formula for the purity $\Pi_{N-m}=\tr(\rho_{N-m}^2)$, where
$\rho_{N-m}$ is the density operator $\rho_N=\ket{\psi}\bra{\psi}$
reduced by $m$ subsystems. We find
\begin{equation} \label{purities}
\Pi_{N-m}=\frac{1 + \gamma^m + \gamma^N + 4 \gamma^{N/2} +
\gamma^{N-m}}{2(1+\gamma^{N/2})^2},
\end{equation}
with $\gamma=1-\epsilon^2=|\braket{\phi_1}{\phi_2}|^2$.

Suppose we create two identical BECs, each in state $\rho_N$, wait
for a time $t_c$ to let their density operators be inelastically
reduced via single particle loss processes to $\rho_{N-m}\otimes
\rho_{N-m'}$, and then let the two BEC's go through a BS like
transformation. As an aside we note that the reduced density
operators emerging from multi particle collisions not only depend
on $\epsilon$, but also on $\ket{\phi_1},\ket{\phi_2}$ and thus
could be used to gain further insight into the properties of the
state. The BS can be implemented either through collisional
interactions between the atoms in two arms of a spatial
interferometer \cite{BS BEC}, or by first turning both BECs into
Mott insulator states \cite{Jaksch98} trapping them in an optical
lattice and then switching on $H_{\rm BS}$. We only consider the
latter method since it corresponds more directly to the situation
of Fig.~\ref{Fig1}. The loss processes which reduce the density
operators $\rho_N$ are stochastic so in general $m \neq m'$ which
means that only $N-n$, where $n=\max\{m,m'\}$, pairs of atoms will
undergo pairwise BS in the lattice. Since only density profiles of
pairs of vertical sites with two atoms contribute to the
interference pattern, measuring the collective density profile
will determine $\tr(\rho_{N-n}^2)$. Plots of different $\Pi_{N-n}$
for an initial number of $N=300$ atoms as a function of $\epsilon$
are presented in Fig.~\ref{Fig2}(b). The dependence of these
curves on $N$ is very weak but for constant $\epsilon$ the values
of $\Pi_{N-n}$ quickly tend towards $1/2$ as $n$ increases.
Therefore from measuring the density profile the determination of
$\epsilon$ from $\Pi_{N-n}$ is best done for small $n\sim 15$. For
a given particle loss rate the average value of $n$ after time
$t_c$ will be known and $\epsilon$ can be found by averaging over
several runs of the experiment performed under identical initial
conditions. We note that this measurement is considerably simpler
than those in the previous entanglement detection schemes, since
we do not require the ability to distinguish between individual
pairs of bosons but only need to find the overall probability of
projecting on the symmetric and antisymmetric subspaces. If the
experimental setup allows to determine the number of pairwise beam
splitters $N-n$, the measurement can be performed in one run and
inelastic processes are not necessary if one can measure the
collective density profile associated with a subset of the pairs
of atoms.

We have presented and investigated a simple quantum network that
detects multipartite entanglement, requiring only two identical
copies of the quantum state and pairwise BS between the
constituents of each copy. We have shown how the network can be
implemented in optical lattices and magnetic microtraps, using
current technology. As examples of its power we have applied the
network to detect entanglement and imperfections in cluster states
and shown that it also can be used to characterize macroscopic
superposition states.

This work was supported by the IRC network on Quantum Information
Processing. C.M.A. thanks Artur Ekert for useful discussions and
is supported by the Funda{\c c}{\~a}o para a Ci{\^e}ncia e
Tecnologia (Portugal).


\begin{thebibliography}{99}
\bibitem{Foot1} When $J \approx U$, an extra phase is introduced during the BS, leading to a different time $T$.
\end{thebibliography}
\end{document}